\documentclass[prl,twocolumn,showpacs,amsmath,amssymb,superscriptaddress,floatfix]{revtex4}
\usepackage{graphicx}
\begin{document}
\title{Coexistence of magnetic fluctuations and superconductivity in the pnictide
high temperature superconductor SmFeAsO$_{1-x}$F$_{x}$ measured by muon
spin rotation} 

\author{A.~J.~Drew} 
\affiliation{University of Fribourg, Department of Physics and
Fribourg Center for Nanomaterials, Chemin du Musee 3, CH-1700
Fribourg, Switzerland} 

\author{F.~L.~Pratt}
\affiliation{ISIS Facility, Rutherford Appleton Laboratory, Chilton,
Oxfordshire OX11 0QX, United Kingdom}

\author{T.~Lancaster}
\affiliation{Oxford University Department of Physics, Clarendon
Laboratory, Oxford OX1 3PU, United Kingdom}

\author{S.~J.~Blundell}
\affiliation{Oxford University Department of Physics, Clarendon
Laboratory, Oxford OX1 3PU, United Kingdom}

\author{P.~J.~Baker}
\affiliation{Oxford University Department of Physics, Clarendon
Laboratory, Oxford OX1 3PU, United Kingdom}

\author{R.~H.~Liu}
\affiliation{Hefei National Laboratory for Physical Sciences at Microscale and Department
of Physics, University of Science and Technology of China, Hefei, Anhui
230026, China}

\author{G.~Wu}
\affiliation{Hefei National Laboratory for Physical Sciences at Microscale and Department
of Physics, University of Science and Technology of China, Hefei, Anhui
230026, China}

\author{X.~H.~Chen}
\affiliation{Hefei National Laboratory for Physical Sciences at Microscale and Department
of Physics, University of Science and Technology of China, Hefei, Anhui
230026, China}

\author{I.~Watanabe}
\affiliation{RIKEN-RAL, Nishina Centre, 2-1 Hirosawa,
  Wako, Saitama, 351-0198 Japan}

\author{V.~K.~Malik}
\affiliation{University of Fribourg, Department of Physics and
Fribourg Center for Nanomaterials, Chemin du Musee 3, CH-1700
Fribourg, Switzerland} 

\author{A.~Dubroka}
\affiliation{University of Fribourg, Department of Physics and
Fribourg Center for Nanomaterials, Chemin du Musee 3, CH-1700
Fribourg, Switzerland} 

\author{K.~W.~Kim}
\affiliation{University of Fribourg, Department of Physics and
Fribourg Center for Nanomaterials, Chemin du Musee 3, CH-1700
Fribourg, Switzerland} 

\author{M.~R\"{o}ssle}
\affiliation{University of Fribourg, Department of Physics and
Fribourg Center for Nanomaterials, Chemin du Musee 3, CH-1700
Fribourg, Switzerland} 

\author{C.~Bernhard} 
\affiliation{University of Fribourg, Department of Physics and
Fribourg Center for Nanomaterials, Chemin du Musee 3, CH-1700
Fribourg, Switzerland} 


\begin{abstract}
Muon-spin rotation experiments were performed on the pnictide high
temperature superconductor SmFeAsO$_{1-x}$F$_{x}$ with x=0.18 and 0.3.
We observed an unusual enhancement of slow spin fluctuations in the vicinity of the superconducting which suggests that the spin fluctuations contribute to the formation of an unconventional superconducting state. 
An estimate of the in-plane penetration depth $\lambda
_{ab}(0)=190(5)$\thinspace nm was obtained, which confirms that
the pnictide superconductors obey an Uemura-style relationship between
$T_{\mathrm{c}}$ and $\lambda _{ab}(0)^{-2}$.
\end{abstract}
\date{\today}
\maketitle

%INTRO\bigskip

The recent discovery of high temperature superconductivity (HTSC) in
the layered tetragonal pnictide compound RFeAsO$_{1-x}$F$_{x}$
(R=La,Nd,Pr,Gd, and Sm) with critical temperatures
$T_{\mathrm{c}}$ above 50 K came as a considerable surprise
\cite{Kamihara1,Chen1,Ren1}. This is the first family of non
copper-oxide-based layered superconductors with $T_{\mathrm{c}}$ exceeding
40\thinspace  K and raises the expectation that even higher $T_{\mathrm{c}}$ values
can be achieved. This discovery also gives rise to the speculation that a common
pairing mechanism is responsible for HTSC\ in cuprates and pnictides.

At first glance, the pnictides appear rather different from the
cuprates. Band structure calculations suggest that they are multiband
superconductors with up to five FeAs-related bands crossing the Fermi-level
\cite{Lebeque1,Haule1,Xu1} as opposed to the cuprates which have only one 
relevant Cu(3d$_{x^{2}-y^{2}}$)O band.
The exchange interaction also seems to be more complex since, besides the
indirect Fe(3d)-As(4p) hybridization, a sizeable direct Fe(3d)-Fe(3d) overlap
has been predicted \cite{Haule1,Xu1} as well as significant frustation 
\cite{Yildirim}. Furthermore, the highest $T_{\mathrm{c}}$ is for electron 
doping rather than the hole doping of the cuprates. 

Nevertheless, there are also some striking similarities, such as HTSC 
emerging on doping away from a magnetically ordered parent compound \cite{Chen2}.  
Neutron measurements on undoped LaFeAsO have revealed commensurate 
spin-density wave (SDW) order of the Fe moments below $T_{\rm N}$=135 K 
with amplitude 0.35$\mu_{\rm  B}$ \cite{Mook1}, a result confirmed by muon %Mandrus1,
spin rotation ($\mu$SR) and M{\"o}ssbauer \cite{Klauss}.  Resistivity 
measurements exhibit an anomaly near $T_{\rm SDW}$ in LaFeAsO and SmFeAsO 
which has been tracked as a function of F-doping. These measurements suggest 
that the magnetic order is rapidly suppressed upon doping and that the 
maximum $T_{\mathrm{c}}$ is achieved just as static magnetic order 
disappears \cite{Chen2}. 
Recent neutron measurements on F-doped superconducting samples confirm this
conjecture since they could not detect any magnetic order \cite{Mook1}.
Thus an important issue is whether weak, slowly fluctuating or strongly
disordered magnetism persists in these superconductors.

%RESULTS\ SHORT

In this letter we report a $\mu $SR study which provides new insight into the magnetic properties of
this new superconductor.  Two polycrystalline samples with nominal
compositions of $x$=0.18 and 0.3 were synthesized by conventional
solid state reaction methods as described in Ref \cite{Chen1,Chen2}. Standard powder x-ray diffraction patterns were measured where all (the main) peaks could be indexed to the tetragonal ZrCuSiAs-type structure for $x$=0.18 ($x$=0.3), as previously reported \cite{Chen1,Chen2}.  DC resistivity and magnetisation measurements were made to determine
$T_{\mathrm{c}}$ ($\Delta T_{\mathrm{c}}$) = 45(3) and 45(4) K for $x$=0.18 and 0.3
corresponding to the midpoint (10\% to 90\% width) of the
resistive and the diamagnetic transitions.

%MUSR\ TECHNIQUE

The $\mu$SR experiments were performed at
the EMU, MuSR and ARGUS instruments of the ISIS facility, Rutherford
Appleton Laboratory, UK, which provides pulsed beams of 100\%
spin polarized muons. 
$\mu $SR measures the time evolution
of the spin polarization of the implanted muon ensemble using the time-resolved asymmetry $A(t)$ of muon decay positrons.
The technique \cite{sjb99,Schenck86} is well
suited to studies of magnetic and superconducting (SC) materials as it
allows a microscopic determination of the internal field distribution
and gives direct access the volume fractions of
SC and magnetic phases \cite{Schenck86}. 
%The $\mu $SR
%technique covers a time window of 10$^{-6}$ to 10$^{-9}$\,s
%and allows one to detect internal magnetic fields as small as
%0.1~G. Spin-polarized positive muons (energy $\approx$ 4.2 MeV,
%lifetime $\approx$ 2.2\,$\mu$s) are
%implanted into the bulk of the sample and stop at well-defined
%interstitial lattice sites which are currently unknown for the pnictides. 
%Each muon spin precesses in the local magnetic
%field $B_{\mu}$ with a precession frequency of $\nu_{\mu}$ =
%($\gamma_{\mu}B_{\mu}/2\pi)$, where $\gamma_{\mu} = 2\pi\times
%135.5$\thinspace  MHz/T is the gyromagnetic ratio of the muon. 

%DATA

Figure 1(a) shows representative spectra of the zero-field (ZF) $\mu
$SR measurements at three different temperatures for $x$=0.18. The relatively fast
relaxation of $A(t)$, which persists even at 200\thinspace K, provides
clear evidence for the presence of sizeable electronic magnetic
moments. We find that these spectra are well described with a
single stretched exponential relaxation function of the form
$A(t)=A(0)G(t)$ where the spin polarization function is $G(t)=\exp[-(\lambda ^{\mathrm{ ZF}}t)^{\beta}]$. This is illustrated in the inset of Fig. 1(a) which shows that the data follow straight lines on a log-log plot of 
$-t/ln(G(t))$ versus $t$ \cite{Heffner01}. The temperature dependences of the relaxation rate,
$\lambda^{\mathrm{ZF}}$, and the exponent, $\beta$, are shown in
Figs.~1(b) and 1(c), respectively. Above 100\thinspace K the
relaxation is exponential with $\beta\approx 1$ and
$\lambda^{\mathrm{ZF}}$ is only weakly temperature dependent. Below
100\thinspace K the value of $\lambda^{\mathrm{ZF}}$ exhibits a
significant increase followed by a saturation below 30\thinspace K
with a low temperature value of $\lambda^{\mathrm{ZF}}\approx 1.2
{\mu}$s$^{-1}$. At the same time $\beta$ decreases continuously
towards $\beta\approx 0.5$.  Notably, the biggest changes occur in the
vicinity of the SC transition at $T_{\mathrm
{c}}=45(3)$\thinspace K, as shown by the vertical dashed line in
Figs~1(b) and (c). The inset of Fig. 1(b) shows corresponding data for the  $x$=0.3 sample where $\lambda^{\mathrm{ZF}}$ is reduced but exhibits a similar temperature dependence.

\begin{figure}
\includegraphics[width=8.2cm]{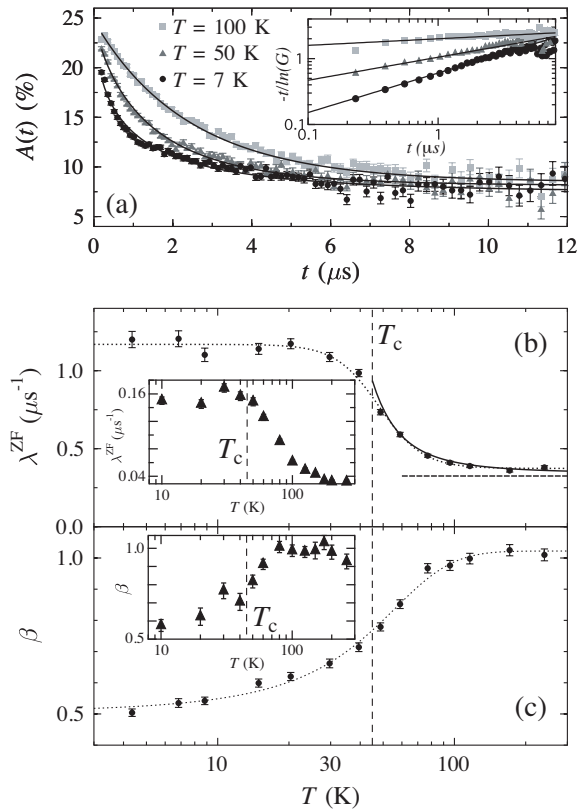}
\caption{(a) Example ZF-$\protect\mu $SR spectra for $x$=0.18. Inset: log-log plot of the polarization function, G(t)  \cite{Heffner01}.   
  (b) Temperature evolution of the fitted relaxation rate $\protect\lambda^{\mathrm
  {ZF}}$ for $x$=0.18. Inset: Corresponding data for $x$=0.3. (c) Shape parameter $\protect\beta $ for $x$=0.18. Inset: Corresponding data for $x$=0.3.
  In (b) the solid line is the
  sum of a temperature-independent component (shown by the horizontal dashed
  line) and an activated component with an activation energy of
  13(2)\thinspace  meV. Dotted lines are a guide to the eye. }
\label{fig1}
\end{figure}

In order to distinguish between static and dynamic
contributions, a longitudinal field (LF) scan was performed at 60\thinspace K
(Fig.\thinspace 2).  We observe an abrupt transition
in the longitudinal
relaxation rate $\lambda^{\mathrm{LF}}$ at around 40\thinspace G but subsequent
increases in LF produce no further significant change.  This unusual behavior cannot be accounted for by a
simple decoupling model for purely static or dynamic spins (dashed
line in Fig.~2) which would
predict $\lambda^{\mathrm{LF}} \propto B^{-2}$ above some critical field
(corresponding either to the internal field in the static case or 
$\nu/\gamma_{\mu}$ in the dynamic case, where $\nu$ is the fluctuation
rate).  

\begin{figure}
\includegraphics[width=8.2cm]{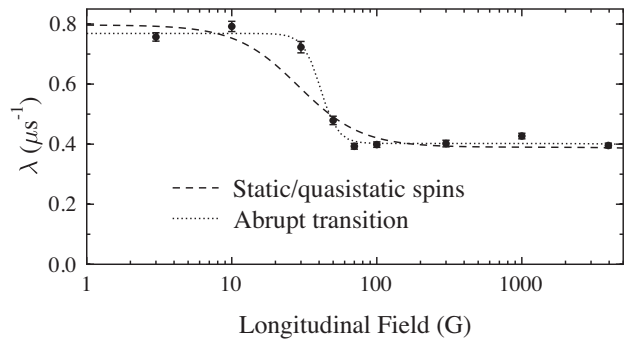}
\caption{Field dependence of the LF relaxation rate
  $\protect\lambda_{\mathrm{LF}}$ at  60\thinspace K for the x=0.18 sample. The dashed line shows a simulation assuming decoupling from a simple static or dynamic local field distribution (see text).  The dotted line is a guide to the eye and illustrates the more abrupt transition which is actually observed.
}
\label{fig2}
\end{figure}

Two potential magnetic sources for the muon relaxation are: (1)
the lanthanide moments in the SmO layers; (2)
magnetic fluctuations originating from spin correlations in the
FeAs layers. The strong doping dependence of the value of $\lambda^{\mathrm{ZF}}$ 
(Fig.\thinspace 1(b)) demonstrates clearly that the observed magnetism cannot be 
explained solely in terms of weakly coupled Sm moments. The data for $T>T_{\rm c}$ 
naturally separates into two components: $T$-independent and $T$-dependent, as indicated by the guides to the eye in Fig. 1(b). The 
$T$-dependent component has an activation energy of 13(2)\thinspace  meV, a typical 
scale for lanthanide moment fluctuations \cite{bewley}.  The amplitude corresponds 
to the non-quenched component seen in LF and we ascribe this component to Sm moments 
which are fluctuating rapidly in this temperature region due to crystal field excitations
\cite{bewley}.  The temperature-independent component which can be
quenched in 40\thinspace G at 60\thinspace K can be identified with 
low energy spin fluctuation processes associated with the FeAs layer.

Although the high temperature relaxation is simple
exponential and can therefore be associated with a single, dominant
fluctuation rate, cooling through $T_{\mathrm{c}}$ results in a reduction in
$\beta$, signifying a range of fluctuation rates and/or local field
amplitudes.  This indicates that the spin dynamics at low temperature
becomes substantially more complex.  Below $T_{\mathrm{c}}$, the activated
behavior ceases and $\lambda^{\mathrm{ZF}}$ saturates, demonstrating that
an additional relaxation channel
becomes dominant. 
Although relaxation with $\beta=0.5$ can be suggestive of glassy
behavior or static disorder, it is unlikely to be the case here.  
Since it is known that the Sm moments order antiferromagnetically below $\sim
3-4$\thinspace K \cite{ding} (and see below), we may preclude the possibility
of the Sm moments forming a static disordered state at a higher temperature.  While it is
possible that static Fe moments could develop in this temperature regime,
the lack of change in $\lambda_{\rm ZF}$ in the region below 30 K where $\beta$
is changing continuously makes this interpretation unlikely.
We can also discard any interpretation of our data which involves a
small fraction of ordered spins, since the observed relaxation
corresponds to the behavior of the overwhelming bulk of the sample
across the entire temperature range; thus the role of any purported
minority phase is not directly probed in these experiments.

Accordingly, our data suggest that the onset of superconductivity in this Sm compound is accompanied or slightly preceeded by the enhancment of slow spin fluctuations which originate (at least partially) in the FeAs layers. This raises the question whether this coincidence is accidential or rather signifies that the spin fluctuations are playing an active role in the SC pairing mechanism. The former scenario is not supported by the similar temperature dependences of the spin fluctuations for the x=0.18 and x=0.3 samples (despite of their different absolute values). Furthermore, the latter scenario is in agreement with the finding that the $T_{\mathrm{c}}$ values for the Sm compounds are almost twice those of the La ones, whose spin fluctuations were shown to be considerably weaker or possibly even absent \cite{Luetkens1,Carlo}. It is also supported by theoretical calculations which show that spin fluctuations emerging in the proximity of an AF or SDW state can mediate or at least significanlty enhance a singlet SC state \cite{Kato88,Machida82}. Notably, the spin fluctuations enhance only unconventional order parameters whereas they suppress conventional ones. 

One might even be tempted to speculate about an unconventional SC state with spin triplet Cooper pairs, similar to (U,Th)Be$_{13}$, UPt$_3$  \cite{Heffner90} and Sr$_2$RuO$_4$ \cite{Luke98} where the beaking of time-reversal symmertry yields spontaneous supercurrents that create internal magnetic fields below $T_{\mathrm{c}}$. However, these magnetic fields should be static rather than dynamic. Also the Sm moment would have to play an important role in this unconventional SC state since no corresponding increase in the ZF relaxation rate has been observed in the related La compound \cite{Luetkens1,Carlo}.  
 
Certainly, our observations call for further investigations to clarify the role of slow spin fluctuations in the SC pairing mechanism and to explore whether the enhanced spin fluctuations in the Sm compound as compared to the La one are brought about by the coupling to the lanthanide moments or rather by the related structural changes \cite{Kamihara1,Chen1}.

%The observation that the steepest changes of  $\lambda_{\rm ZF}$ occur 
%near $T_{\mathrm{c}}$ together with the indications from transverse field 
%measurements (as shown below) that precursor SC correlations may exist up 
%to 70~K suggest that this anomalous behavior may be related to superconductivity, 
%i.e. that the transition into the SC state might be associated with 
%an enhancement of the low energy spin fluctuations of the Fe moments
%and a related change in the Sm moment fluctuations. 

%Future experiments on samples with lower F doping
%will tell us whether this transition scales indeed with $T_{\mathrm{c}}$ or whether
%it is rather linked with a precursor of the SDW state that has been directly observed
%in the undoped system. 

%One might even be tempted to speculate that an unconventional SC state with spin triplet Cooper pairs could be realised in these pnictides similar as in UPt3 and Sr2RuO4 where spontaneous supercurrents develop below Tc and give rise to an increase of the muSR relaxation rate. However, such an interpretation does not appear to be consistent with the muSR data on the La compound where no corresponding increase of the ZF-mSR relaxation rate in the vicinity of Tc was observed. In addition the magnetic correlations in the Sm compound appear to be dynamic in nature rather then static as would be expected for such aa triplet scenario. 

\begin{figure}
\includegraphics[width=8.2cm]{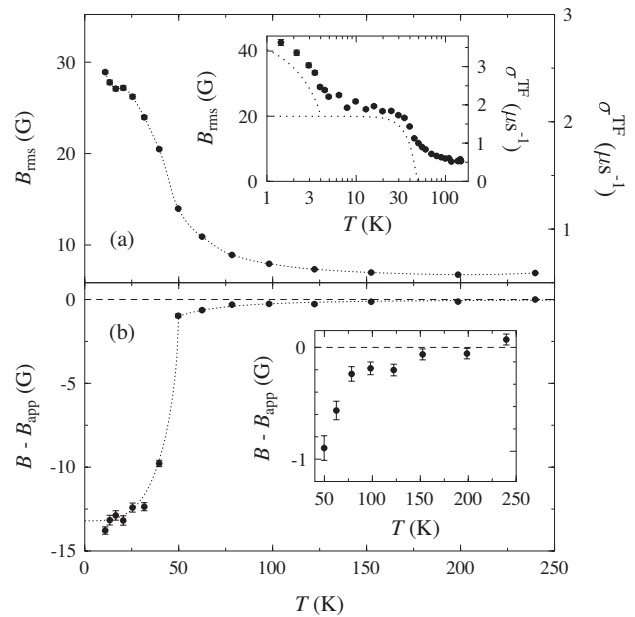}
\caption{ (a) The main component  of the TF relaxation rate $\sigma^{\mathrm{TF}}$, taken on the ARGUS spectrometer.
  The inset shows in addition data below 4 K from the MuSR spectrometer, which reveals a steep
increase of $\protect\sigma ^{\mathrm{TF}}$, likely due to the ordering of
the Sm moments.  The dashed lines illustrate the respective contributions from the superconducting vortex lattice and the Sm ordering. (b) The diamagnetic shift due to the development of the superconducting  state.  The inset shows a detail of the behavior above $T_{\rm
  c}$. Dotted lines are guides to the eye.}
\label{fig3}
\end{figure}

Unconventional superconductivity is also supported by our transverse field TF-$\mu $SR measurements which yield a low temperature value of the in-plane magnetic penetration depth, $\lambda_{ab}(0)$,  that falls close to the so-called Uemura-line of the cuprate HTSC \cite{Uemura89}. 
The TF-$\mu$SR\ spectra for $B_{\mathrm{app}}$ = 100\thinspace G
were well described with a sum of two Gaussian functions using the form 
\begin{equation}
A(t)=\sum_{i=1,2}A_{i}(0)\,\cos(\gamma_{\mu} B_{i} t) \exp\left[-{\left(\frac{%
\sigma_{i} ^{\mathrm{TF}} t}{2}\right)^2}\right],
\end{equation}
 where $A_{i}$, $B_{i}$, and
$\sigma_{i}^{\mathrm{TF}}$ correspond to the amplitude, the local magnetic
field at the muon site, and the relaxation rate, respectively.  The
second weakly-damped component reflects the small background from muons not stopping in the sample.  The first,
dominant component is due to the sample, and its temperature dependence is shown in
Fig.\thinspace 3.
A sharp rise of $\sigma^{\mathrm{TF}}$ [Fig. 3(a)] is seen below
$T_{\mathrm{c}}$ which exceeds that which would be expected from the ZF
data; this additional contribution reflects the formation of the
vortex lattice.  This interpretation is confirmed by
an observed diamagnetic shift of $\sim 13$\thinspace G [plotted in Fig.~3(b)]
which also occurs at $T_{\mathrm{c}}$.  The
additional steep rise of $\sigma^{\mathrm{TF}}$ below
4 K [see inset to Fig.~3(a)] likely represents additional
local-field broadening due to the ordering of the Sm moments.

\begin{figure}
\includegraphics[width=8.2cm]{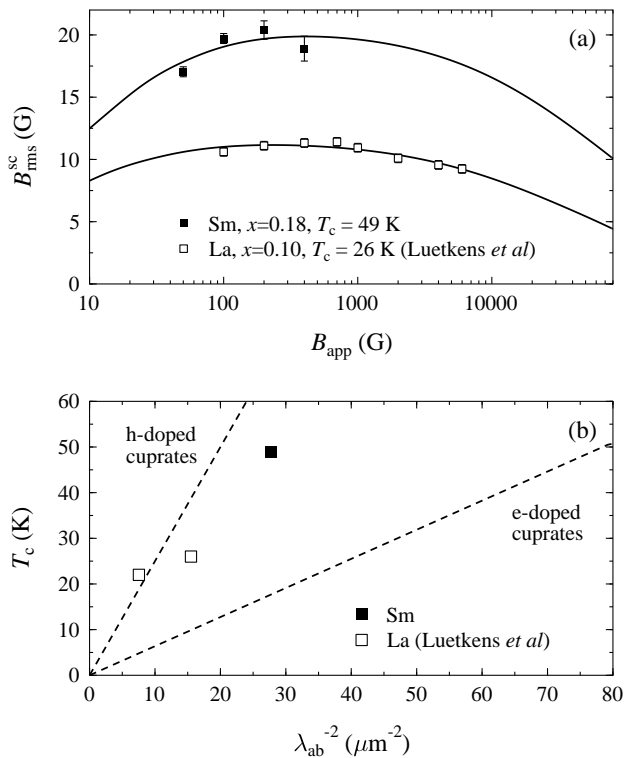}
\caption{(a) The field dependence of the superconducting 
vortex contribution to the rms linewidth of our sample.  The fitted field
  dependence is given by a powder averaged version of the model in Ref.~\cite{brandt}. 
  Data for the La compound from Ref.~\cite{Luetkens1} are shown for comparison \cite{Luetkens2,kappa} (b).  The Uemura plot for the pnictide superconductors measured to date by $\mu$SR. Trend lines for the cuprates are shown for comparison.}
\label{fig4}
\end{figure}

Notably, on cooling towards $T_{\mathrm{c}}$ there is a gradual
onset of a diamagnetic shift already at about 70~K, which is detailed in the inset to Fig.3(b).
As was already noted above, this result could be an indication for a precursor SC state with
a transition temperature higher than the bulk $T_{\mathrm{c}}$ or the onset of SC fluctuations
above $T_{\mathrm{c}}$. However, at present we cannot rule out the possibility that the slowing down of the spin fluctuations
leads to this reduction of the local field.
In any case, the sharp onset of the diamagnetic shift at $T_{\mathrm{ c}}$
and the corresponding increase in $\sigma^{\mathrm{TF}}$ allows us to
provide an estimate of the in-plane magnetic penetration depth.
Allowing for an additional root-exponential damping in the dominant term of (1) to take account of the contribution to the relaxation from magnetic fluctuations (which is known from the ZF measurements), the SC vortex contribution to the total linewidth $B^{\rm sc}_{\rm rms}$ is obtained.
This is plotted in Fig.~4(a) against applied field at 10 K and from this data 
we derive $\lambda_{ab} = 190(5)$\thinspace nm.    
This estimate is
derived from fitting to the numerical results of a recent detailed Ginzburg-Landau vortex
lattice calculation \cite{brandt}, taking a polycrystalline average for our powder sample in the high anisotropy limit, under the assumption that the length scales $\lambda$ and $\xi$ diverge following $1/ \cos\theta$ as the field orientation approaches the plane at $\theta$ = 0.

Since the estimate is made at 0.2~$T_{\mathrm{c}}$, it should provide a good account of $\lambda_{ab}(0)$, assuming a two-fluid type of saturating temperature dependence. 
Note that we are unable to establish whether any additional gap-node related linear term might be present at low temperatures due to the extra relaxation contribution below 4 K.
 Our value of $\lambda_{ab}$ is shorter than those found
for LaFeAsO$_{1-x}$F$_x$ by Luetkens {\sl et al.}  (254(2)\thinspace nm for
$x=0.1$ and 364(8)\thinspace nm for $x=0.07$ \cite{Luetkens1,Luetkens2}), reflecting the higher
$T_{\rm c}$ of our compound and hence larger superfluid stiffness
(proportional to $\lambda_{ab}^{-2}$).
Fig.~4(b) shows the Uemura plot \cite{Uemura89} for the high temperature pnictide superconductors measured to date by $\mu$SR.  
It appears that the overall trend lies closer to that of the hole-doped than the electron-doped cuprates. 

In conclusion, the $\mu$SR results on polycrystalline
SmFeAsO$_{1-x}$F$_{x}$ with $x$=0.18 and 0.3 provide clear evidence for the coexistence
and interplay of superconductivity and dynamic magnetic correlations. 
The magnetic correlations exhibit a complex temperature dependence and a significant contribution of magnetic fluctuations to the enhanced $T_{\rm c}$ is suggested. 
From TF measurements we obtained an estimate
of the in-plane magnetic penetration depth of $\lambda
_{ab}=190(5)$\thinspace nm, which comes rather close to the Uemura
line of the hole doped cuprates.

This work is supported by a Schweizer Nationalfonds (SNF) grant
200020-119784, a Deutsche Forschungsgemeinschaft (DFG) grant BE2684/1-3 in FOR538 and the UK EPSRC. We acknowledge helpful discussions with D. Baeriswyl, A. T. Boothroyd, Ch. Niedermayer and M. Siegrist and numerical data from E.H.~Brandt.

{\em Note added:} Since submission of this Letter we learned of Ref. \cite{Khasanov} which interprets the ZF-$\mu$SR of SmFeAsO$_{0.85}$ in terms of two exponential components and attributes the relaxation solely to weakly coupled Sm moments. Such an interpretation is inconsistent with our relaxation data and its LF and x dependences.


\begin{thebibliography}{99}
\bibitem{Luetkens2}
Using the same
analysis procedure on the data for the 
$x=0.1$ sample of Ref.~\onlinecite{Luetkens1} yields
$\lambda_{ab}=254(2)$\thinspace nm and $\kappa = 89(10)$ (Fig.4a), fully 
consistent with the $\lambda_{ab}$ estimate of these authors, however
our analysis shows that the field dependence of the linewidth is well 
described by a standard Ginzburg-Landau model (Ref.16).
\bibitem{kappa}
Due to the more limited field range of the Sm data, the $\kappa$ estimate for La was used in the Sm analysis.
\end{thebibliography}
\end{document}